\documentclass[11pt,a4paper]{article}
\pdfoutput=1
\usepackage{jcappub}
\graphicspath{{Fig/}}
\title{Evolution of a simple inhomogeneous anisotropic cosmological model with diffusion}

\author{Dmitry Shogin}
\author{and Sigbj{\o}rn Hervik}
\affiliation{Faculty of Science and Technology, University of Stavanger, N-4036 Stavanger, Norway}

\emailAdd{dmitry.shogin@uis.no}
\emailAdd{sigbjorn.hervik@uis.no}


\abstract{
We investigate a simple inhomogeneous anisotropic cosmology (plane symmetric~$G_2$ model) filled with a tilted perfect fluid undergoing
velocity diffusion on a scalar field. Considered are two types of fluid: dust and radiation. We solve the system of Einstein field equations and diffusion equations numerically and demonstrate how the universe evolves towards its future asymptotic state. Also, typical time scales of characteristic processes are determined. The obtained results for dust- and radiation-filled cosmologies are compared to each other and to those in the diffusionless case, giving a hint on which effects can be the result of including diffusion in more complicated models. For example, diffusion causes the accelerated expansion stage to arrive at later times.}
\begin{document}
\maketitle 
\flushbottom

\section{Introduction}
\label{sec:introduction}
Diffusion forces cause some important physical phenomena, such as heat conduction and Brownian motion. During recent years a lot of attention has been paid to investigating diffusion effects also in special and general relativity, e.\thinspace g. \cite{Dudley1966, Risken1996, Rendall2004, Haba2009, Herrmann2009, Herrmann2010, Calogero2011} (see~\cite{Herrmann2009, Herrmann2010, Calogero2011} for further references). It is supposed that diffusion may play a fundamental role not only at microscopic scales, but also in the large-scale dynamics of the matter in the universe. In the cosmological case, the particles of a fluid matter can be represented by galaxies in space, and the role of a background medium can be played, for example, by a scalar field \cite{Calogero2011, Calogero2012}.

\par 
In \cite{Calogero2012}, a simple homogeneous and isotropic FRW-model of the universe is investigated. The universe is treated to be filled with a perfect fluid interacting with a scalar field background. The perfect fluid is undergoing velocity diffusion on the scalar field. It was shown that the presense of diffusion can drastically affect the dynamics even in this simplest cosmology.
\par 
In the current paper, we are investigating the diffusion effects in plane symmetric~$G_2$ cosmologies \cite{Wainwright1997, Ellis2012}. The motivation behind this choice is that the given models possess important specific properties (such as inhomogeneity and anisotropy), but at the same time they are not too complicated from a mathematical point of view. Following \cite{Calogero2012}, we consider interaction between a perfect fluid and a scalar field background, but in contrast to \cite{Calogero2012}, the fluid is tilted (the four-velocity of the fluid is not in general aligned to that of fundamental observers \cite{King1973}). 
\par 
The paper is organized as follows. In Sec.\thinspace \ref{sec:the_diffusion_model} a brief description of the diffusion model is given. In Sec.\thinspace \ref{sec:geometry_of_the_model} we discuss the geometry of spacetime in the considered cosmological model. Sec.\thinspace \ref{sec:energy-matter_content} is devoted to discussing the energy-matter content of the given cosmology. In Sec.\thinspace \ref{sec:dimensionless_variables} we introduce the scale-independent dimensionless variables and do the gauge-fixing, after which the complete system of equations governing the dynamics of the universe is derived in Sec.\thinspace \ref{sec:normalized_equations}. The choice of initial and boundary conditions is discussed in Sec.\thinspace \ref{sec:ibc}. The results of numerical simulation are presented in Sec.\thinspace \ref{sec:dynamics} and thouroughly analyzed in Sec.\thinspace \ref{sec:analysis}, where also some conclusions are drawn. Summary is presented in Sec.\thinspace \ref{sec:summary}.

\section{The diffusion model}
\label{sec:the_diffusion_model}
\par 
The mathematical and physical background for a simple cosmological diffusion model is given in \cite{Calogero2011}. Here we will outline the main idea.
The spacetime geometry is governed by the Einstein field equations
\begin{equation}
R_{ab}-\frac{1}{2}Rg_{ab}=\mathcal{T}_{ab},
\end{equation}
where the Bianchi identities impose the restriction 
\begin{equation}
\nabla_a\mathcal{T}^{ab}=0
\end{equation}
Since the kinetic energy of particles is not conserved under the action of diffusion forces, the energy-momentum tensor~$T_{ab}$ describing such matter fails to be divergence-free. Therefore,~$T_{ab}$ cannot enter the field equations alone, and the energy-momentum tensor~$\tilde{T}_{ab}$ of the background medium should also be included in the equations. The combination of these two tensors~$\mathcal{T}_{ab}$ meets the mentioned requirement:
\begin{align}
& \mathcal{T}_{ab}=T_{ab}+\tilde{T}_{ab}, \\
& \text{with } \nabla_a T^{ab}=-\nabla_a \tilde{T}^{ab}. 
\end{align}
In one of the simplest cases \cite{Calogero2011, Calogero2012}, a perfect fluid matter is undergoing velocity diffusion on a cosmological scalar field~$\phi$ (with~$\tilde{T}_{ab}=-\phi g_{ab}$), which is playing the role of the background. The modified field equations can be written as
\begin{equation}
\label{eq:EFE_modified}
R_{ab}-\frac{1}{2}g_{ab}+\phi g_{ab}=T_{ab},
\end{equation}
the diffusion equations being
\begin{align}
\label{eq:diffusion_equations}
& \nabla_a T^{ab}=\nabla_a (\phi g^{ab})=DJ^b, \\
& J^a=n \hat{u}^a, \\
& \nabla_aJ^a=0.
\end{align}
Here $D$ is a positive dimensionfull constant of diffusion,~$J^a$ is the conserved current density of the fluid, $n$~is the number of matter particles per unit volume (concentration or number density), and~$\hat{u}^a$ is the~4-velocity of the fluid defined by the fluid flow.
\par 
Note that scalar potential represents vacuum energy and enters the field equations in the same manner as the cosmological constant. Therefore,~$\phi$ can be treated as a generalization of cosmological constant. However, one should note the difference between models with scalar potential and with a "variable cosmological constant": in the former case the dynamics of potential is governed by the diffusion equations~(\ref{eq:EFE_modified}) and~(\ref{eq:diffusion_equations}), while in the latter case it is prescribed a priori \cite{Calogero2011}.
\par 
In vacuum~($T_{ab}=0$) or in the absence of diffusion~($D=0$), potential~$\phi$ becomes "ordinary" cosmological constant.

\section {The geometry of the model}
\label{sec:geometry_of_the_model}

We are looking for solutions of (\ref{eq:EFE_modified}), corresponding to a plane symmetric $G_2$ cosmology with diagonal metric:
\begin{equation}
ds^2=-T^2(t,x)dt^2+F^2(t,x)dx^2+G^2(t,x)[dy^2+dz^2].
\end{equation}
We shall use the orthonormal frame approach \cite{Wainwright1997} and introduce the following linearly independent vector operators:
\begin{equation}
\label{def:vector_operators}
\begin{split}
\mathbf{e}_0 &= \frac{1}{T(t,x)}\partial_t, \qquad \mathbf{e}_2 = \frac{1}{G(t,x)}\partial_y, \\
\mathbf{e}_1 &= \frac{1}{F(t,x)}\partial_x, \qquad \mathbf{e}_3 = \frac{1}{G(t,x)}\partial_z.
\end{split}
\end{equation} 
These operators can also be written in terms of frame coefficients:
\begin{equation}
\mathbf{e}_0=M^{-1}\partial_t, \qquad \mathbf{e}_\alpha=e_\alpha^\alpha\partial_{x^\alpha}.
\end{equation}
The state vector $\mathbf{x}_{\rm G}$ of the gravitational field variables reduces to
\begin{equation}
\label{def:grav_field_variables}
\mathbf{x}_{\rm G} = (H, \sigma_{\alpha \beta}, \dot{u}_\alpha, a_\alpha),
\end{equation}
with
\begin{align}
& \sigma_{\alpha \beta} = {\rm diag}(\sigma_{11}, \sigma_{22},\sigma_{33}), \quad \sigma_{22}=\sigma_{33}=-\frac{1}{2}\sigma_{11}, \\ 
& \dot{u}_\alpha=(u_1,0,0), \\
& a_\alpha=(a_1,0,0).
\end{align}
The variables have following physical and geometrical sense (see \cite{Lim2004a} for details):~$H$~is the Hubble scalar,~$\sigma_{\alpha \beta}$~is the rate of shear tensor,~$\dot{u}_\alpha$~is the acceleration vector, and~$a_\alpha$~is the spatial curvature variable (set further to zero by gauge choice).
\par 
Now the field equations~(\ref{eq:EFE_modified}) can be written in terms of vector operators~(\ref{def:vector_operators}) and gravitational field variables~(\ref{def:grav_field_variables}).

\section {Energy-matter content of the model}
\label{sec:energy-matter_content}

The energy-matter content of the model is represented by a tilted perfect fluid with energy-momentum tensor~$T_{ab}$ and a scalar field given by potential~$\phi$. The fluid undergoes velocity diffusion on the scalar field, which is described by the diffusion equations (\ref{eq:diffusion_equations}).
\par 
In the frame comoving with the fluid the energy-momentum tensor has the perfect fluid form:
\begin{equation}
T_{ab}=\hat{\rho}\hat{u}_a\hat{u}_b+\hat{p}(g_{ab}+\hat{u}_a\hat{u}_b),
\end{equation}
where $\hat{\rho}$ and $\hat{p}$ are respectively the fluid density and pressure measured in the comoving frame; $\hat{u}_a$ is the 4-velocity of the fluid, with $\hat{u}_a\hat{u}^a=-1.$
We consider a fluid obeying the barotropic equation of state:
\begin{equation}
\hat{p}(\hat{\rho})=(\gamma-1)\hat{\rho},
\end{equation}
$\gamma$ being a constant parameter in the range $1<\gamma<2$.
\par 
In the original orthonormal frame the energy-momentum tensor takes the imperfect fluid form and can be decomposed with respect to a unit timelike vector~$\mathbf{u}$:
\begin{equation}
T_{ab}=\rho u_a u_b + 2q_{(a}u_{b)}+(g_{ab}+u_a u_b)p+\pi_{ab}, 
\end{equation}
which in the current case reduces to
\begin{equation}
\begin{split}
T_{00} &= \rho, \\
T_{01} &= -q_1, \\
T_{\alpha \beta} &= {\rm diag}(p+\pi_{11}, p+\pi_{22}, p+\pi_{33}),
\end{split}
\end{equation}
and the state vector~$\mathbf{x}_{\rm M}$ of the matter variables becomes
\begin{equation}
\mathbf{x}_{\rm M}=(\rho, q_\alpha, p, \pi_{\alpha \beta}),
\end{equation}
with
\begin{align}
& \pi_{\alpha \beta} = {\rm diag}(\pi_{11}, \pi_{22},\pi_{33}), \quad \pi_{22}=\pi_{33}=-\frac{1}{2}\pi_{11}, \\ 
& q_\alpha=(q_1,0,0).
\end{align}
The variables have the following physical sense:~$\rho$~is the energy density,~$q_\alpha$~is the energy flux density,~$p$~is the isotropic pressure, and~$\pi_{\alpha \beta}$~is the anisotropic pressure tensor.

The connection between $\mathbf{\hat u}$ and $\mathbf{u}$ can be expressed by
\begin{equation}
\hat{u}^a=\Gamma(u^a+v^a),
\end{equation}
where $v^a=(0,v^\alpha)$ and $\Gamma=(1-v_\alpha v^\alpha)^{-\frac{1}{2}}$. In our case the tilt of the fluid (its three-velocity) has only one degree of freedom:~{$v^\alpha=(V,0,0)$. Now the matter variables can be expressed in terms of energy density and velocity:
\begin{align}
p  		&= \frac{(\gamma-1)(1-V^2)+\frac{1}{3}\gamma V^2}{G_+}\rho, \\
q_1 		&= \frac{\gamma V}{G_+}\rho, \\
\pi_{11} &= \frac{2}{3}\frac{\gamma V^2}{G_+}\rho,
\end{align} 
where a new function~$G_+$ was introduced by
\begin{equation}
G_+=1+(\gamma-1)V^2.
\end{equation}

\section{Dimensionless variables and Gauge choice}
\label{sec:dimensionless_variables}

Dimensionfull and scale-dependent variables typically diverge when approaching a singularity, and so it is important to write the equations in terms of dimensionless and scale-invariant quantities.
\par 
The area expansion of the $G_2$ orbits is given by 
\begin{equation*}
\Theta_{AB}=H\delta_{AB}+\sigma_{AB},
\end{equation*}
and the average expansion rate is thus
\begin{equation}
\beta = \frac{1}{2}{\Theta_C}^C=H+\sigma_{22}=H-\frac{1}{2}\sigma_{11}.
\end{equation}
This parameter~$\beta$ has the same dimension as the Hubble rate~$H$ and can be used as a normalization factor \cite{Wainwright1997}. The normalized variables are defined as follows:
\begin{equation}
\begin{split}
(\Sigma_{\alpha \alpha},\dot{U},A)=\mathbf{X}_{\rm G}  &= (\sigma_{\alpha \alpha},\dot{u}_1,a_1)/\beta, \\
(\Omega,Q_1,P,\Pi_{\alpha \alpha},\Phi)=\mathbf{X}_{\rm M} &= (\rho,q_1,p,\pi_{\alpha\alpha},\phi)/3\beta^2, \\
\mathcal{N}																  &= Dn/3\beta^3.
\end{split}
\end{equation}
Note that the constant of diffusion~$D$ is now encapsulated in the diffusion term~$\mathcal{N}$ and will not appear in the normalized equations.
\par 
The new differential operators are introduced by
\begin{equation}
\begin{split}
\boldsymbol{\partial}_0      &= \frac{1}{\beta} \mathbf{e}_0      =\mathcal{M}^{-1}\partial_t, \\
\boldsymbol{\partial}_\alpha &= \frac{1}{\beta} \mathbf{e}_\alpha =E_\alpha^\alpha \partial_{x^\alpha},
\end{split}
\end{equation}
where the scale-invariant frame coefficients satisfy
\begin{equation}
\mathcal{M}=M\beta, \qquad E_\alpha^\alpha=\frac{1}{\beta}e_\alpha^\alpha.
\end{equation}

\par 
We define deceleration parameter~$q$ and~$\beta$-gradient by
\begin{align}
q+1 &= -\frac{\boldsymbol{\partial}_0 \beta}{\beta}=-\frac{1}{\beta^2}\mathbf{e}_0(\beta), \\
  r &= -\frac{\boldsymbol{\partial}_1 \beta}{\beta}=-\frac{1}{\beta^2}\mathbf{e}_1(\beta). 
\end{align}
Using $\beta$-normalization is convenient together with an appropriate gauge choice. We use the separable area gauge and timelike area gauge \cite{Lim2004a}, given respectively by 
\begin{align}
\dot{U} 	&= r,\\
A			&=	0,
\end{align}
enabling us to reparametrize time variable to set~$\mathcal{M}=1$ and~$\boldsymbol{\partial}_0=\partial_t$.
It is also convenient to replace~$\Sigma_{11}$ by
\begin{equation}
\Sigma_{11}=-2\Sigma.
\end{equation}

\section{Normalized equations}
\label{sec:normalized_equations}

After normalization we obtain the following system of equations:

{\sc The field equations}
\begin{align}
0					  &= 1-2\Sigma-\Omega-\Phi, \label{eq:normalized_equaions:Omega_Constraint} \\
r					  &= -\frac{3}{2}\frac{\gamma V}{G_+} \Omega, \\
q					  &= \frac{1}{2} \Bigl[1+3\frac{(\gamma-1)(1-V^2)+\gamma V^2}{G_+}\Omega -3\Phi  \Bigr], \label{eq:normalized_equaions:q_Constraint}\\
\partial_t\Sigma &= -(q+3\Sigma)(1-\Sigma)+2\Sigma +\frac{(3\gamma-2)+(2-\gamma)V^2}{2G_+}\Omega-\Phi-\frac{1}{3}\boldsymbol{\partial}_1 r.
\end{align}

{\sc The frame coefficient relations}

\begin{align}
\partial_t E_1^1					&= (q+3\Sigma)E_1^1, \label{eq:normalized_equaions:E11_Evolution} \\
\partial_t E_A^A 					&= q E_A^A, \label{eq:normalized_equaions:EAA_Evolution}\\
\boldsymbol{\partial}_1 E_A^A &= r E_A^A. \label{eq:normalized_equaions:EAA_Constraint}
\end{align}

{\sc The diffusion equations}

\begin{align}
\partial_t \Phi  	&= 2(q+1)\Phi-\Gamma \mathcal{N}, \label{eq:normalized_equaions:Potential_Evolution}\\
\boldsymbol{\partial}_1\Phi &= 2r\Phi+ V\Gamma \mathcal{N}, \label{eq:normalized_equaions:Potential_Constraint} \\
\partial_t\Omega &= -\frac{\gamma V}{G_+}\boldsymbol{\partial}_1\Omega+\frac{\gamma \Omega}{G_+}\Biggl[2\frac{G_+}{\gamma}(q+1)-3(1-\Sigma)(1+\frac{1}{3}V^2)+2\Sigma V^2 \nonumber \\
					  & \qquad -\boldsymbol{\partial}_1V+V\boldsymbol{\partial}_1\ln{G_+}\Biggr]+\Gamma \mathcal{N}, \label{eq:normalized_equaions:Omega_Evolution}\\
\partial_tV		  &= -V \boldsymbol{\partial}_1V+\boldsymbol{\partial}_1G_+-\frac{\gamma-1}{\gamma}(1-V^2)(\boldsymbol{\partial}_1\ln\Omega-2r)-r \nonumber \\
					  &\qquad +(\overline{M}+2\Sigma)V-\frac{\gamma-1}{\gamma}\frac{VG_+}{\Gamma G_-}\frac{\mathcal{N}}{\Omega}, \label{eq:normalized_equaions:Velocity_Evolution} \\
\partial_t \mathcal{N}		  &= -V\boldsymbol{\partial}_1\mathcal{N}+ \frac{\gamma-1}{\gamma}\frac{\Gamma V^2 G_+}{G_-}\frac{\mathcal{N}^2}{\Omega}-\mathcal{N}V\Gamma^2\Biggl[\boldsymbol{\partial}_1G_+ \nonumber \\ 
					  &\qquad -\frac{\gamma-1}{\gamma}(1-V^2)(\boldsymbol{\partial}_1\ln\Omega-2r)-r+(\overline{M}+2\Sigma)V\Biggr]  \nonumber \\
					  &\qquad +\mathcal{N}\Biggl[2rV+3(q+\Sigma)-\boldsymbol{\partial}_1V\Biggr], \label{eq:normalized_equaions:N_Evolution}  
\end{align}
where
\begin{align}
\overline{M}			  &= \frac{1}{G_-}\Biggl[(\gamma-1)(1-V^2)\boldsymbol{\partial}_1V-(2-\gamma)V\boldsymbol{\partial}_1\ln{G_+} \nonumber \\
					  &\qquad +\frac{\gamma-1}{\gamma}(2-\gamma)(1-V^2)V(\boldsymbol{\partial}_1\ln{\Omega}-2r)+G_-rV \nonumber \\
					  &\qquad +(3\gamma-4)(1-V^2)(1-\Sigma)-2(2-\gamma)\Sigma V^2 \Biggr], \\
G_{\pm}			  &=1 \pm (\gamma-1)V^2, \\
\Gamma			  &=  (1-V^2)^{-\frac{1}{2}}.
\end{align}

Comparing the given system to one obtained in \cite{Lim2004a}, we can see that equations~{(\ref{eq:normalized_equaions:Omega_Constraint})--(\ref{eq:normalized_equaions:EAA_Constraint})} have the same form,~(\ref{eq:normalized_equaions:Potential_Evolution})--(\ref{eq:normalized_equaions:Velocity_Evolution}) are modified by presence of the diffusion terms, and equation~(\ref{eq:normalized_equaions:N_Evolution}) is completely new. 
\par 
Note also that equations~(\ref{eq:normalized_equaions:EAA_Evolution}) and~(\ref{eq:normalized_equaions:EAA_Constraint}) decouple from the rest of the system.

The de Sitter equilibrium points for the system of EFEs and Diffusion equations are given by
\begin{equation}
\label{def:deSitter}
\begin{split}
& r=\Sigma=\Omega=\mathcal{N}=0, \\
& \Phi =1, \quad q=-1, \\
& \left\{ \begin{array}{ll} V=0 \text{~or~} V=\pm 1, & \text{if}~\gamma=1, \\
 V=V(x), & \text{if}~\gamma \neq 1. \end{array} \right. 
\end{split}
\end{equation}
Another set of equilibrium points is the Robertson-Walker state:
\begin{equation}
\begin{split}
& r=\Sigma=\Phi=\mathcal{N}=V=0, \\
& \Omega=1, \quad q=\frac{3}{2}\gamma-1.
\end{split}
\end{equation}
Elementary stability analysis shows that the de Sitter state is stable, while the Robertson-Walker state is not. We are particularly interested in solutions which evolve from the state close to Robertson-Walker to de Sitter. This corresponds to the behaviour of the standard model (see e.\thinspace g. \cite{Groen2007}), confirmed by observational data from WMAP and Planck spacecraft. We treat the two most important cases: a model filled with~$\gamma=1$ (dust) and~$\gamma=4/3$ (radiation).

\section{Initial and boundary conditions}
\label{sec:ibc}

The set of variables in the system is:
\begin{equation}
\label{def:variables}
\mathbf{Y}=(E_1^1;\Omega;\Phi;\mathcal{N};V;r;q;\Sigma).
\end{equation}
For numerical simulation we use the system of 5 evolution PDEs~(\ref{eq:normalized_equaions:E11_Evolution}),~(\ref{eq:normalized_equaions:Potential_Evolution}),~(\ref{eq:normalized_equaions:Omega_Evolution})--(\ref{eq:normalized_equaions:N_Evolution}) and 3 algebraic relations~(\ref{eq:normalized_equaions:Omega_Constraint})--(\ref{eq:normalized_equaions:q_Constraint}). In addition, there is a constraint~(\ref{eq:normalized_equaions:Potential_Constraint}), which is taken for error control. We consider solutions satisfying $2\pi$-periodic initial and boundary conditions:
\begin{equation}
\begin{split}
\mathbf{Y}(t,x)	&= \mathbf{Y}(t,x+2\pi),\\
E_1^1(0,x)			&= E_0, \\
\mathcal{N}(0,x)	&= \mathcal{N}_0-\epsilon \mathcal{N}_1 \sin x, \\
\Phi(0,x)			&= \Phi_0-\epsilon \Phi_1 \sin x, \\
\Omega(0,x)			&= \Omega_0+\epsilon \Omega_1 \sin x, \\
(\text{if}~\gamma=1)~~\quad V(0,x) 				&= \epsilon \cos x, \\
(\text{if}~\gamma = 4/3)~~~\quad r(0,x)				&= -2\epsilon \Omega_0 \cos x, 
\end{split}
\end{equation}
where~$\epsilon$~is a small parameter,~$~E_0,~\mathcal{N}_0,~\Phi_0,~$and$~\Omega_0$ are the space-average initial values (close to Robertson-Walker) of the corresponding physical quantities, while~$\mathcal{N}_1,~\Phi_1,$~and~$\Omega_1$ are constants describing spatial inhomogenity. The following values are considered:
\begin{align*}
\Omega_0			& = 0.70 \dots 0.90, \\
\Phi_0			& = 0.10 \dots 0.30, \\
\mathcal{N}_0	& = 0.10 \dots 0.20, \\
\Omega_1			&= \Phi_1 = \mathcal{N}_1 = 0.5, \\
E_0				& = 0.1, \quad \epsilon	= 0.01.
\end{align*}
If we assume the velocity to be small during all the evolution process, the original system can be linearized with respect to~$V$ and its derivatives. Constraints show that this procedure does not significantly affect the errors, which do not exceed~$6\cdot 10^{-3}$ (dust) and~$10^{-2}$ (radiation) in the beginning of simulation, rapidly decrease and tend to zero.

\section{Dynamics of the variables}
\label{sec:dynamics}

It can be easily seen from the equations~(\ref{eq:normalized_equaions:Omega_Constraint})--(\ref{eq:normalized_equaions:N_Evolution}) that it takes infinite time for the variables~(\ref{def:variables}) to reach their equilibrium values~(\ref{def:deSitter}). However, we can consider time points where the difference between the current value of a variable and its equilibrium value drops below a certain level and becomes small. In this paper we assume that quantity~$Y_k$ has approached its equilibrium value, if the following condition is satisfied:
\begin{equation}
\label{def:approaching}
\vert Y_k(t,x)-Y_k(+\infty,x) \vert < \epsilon \vert Y_k(0,x) \vert. 
\end{equation}
So, the quantity~$\epsilon$ is also used as "sensitivity threshold" for the variables of the system.
\par 
We start the simulation at~$t=0$. According to imposed initial conditions, the cosmology is close to the Robertson-Walker state at this timepoint.
\par 
The simulation is stopped at the moment~$t=\tau_{SB}$, when the system comes close in the sense of (\ref{def:approaching}) to the silent boundary (far asymptotic future \cite{Lim2004}). At the silent boundary the frame coefficient~$E_1^1$ vanishes and solution becomes spatially homogeneous.
\par
The transition stage ends when physical quantities~$\Omega$,~$\Phi$, and~$q$ approach their de Sitter values. We denote this point by~$t=\tau$.

\begin{figure}[t]
\begin{center}
\includegraphics[width=0.7\linewidth]{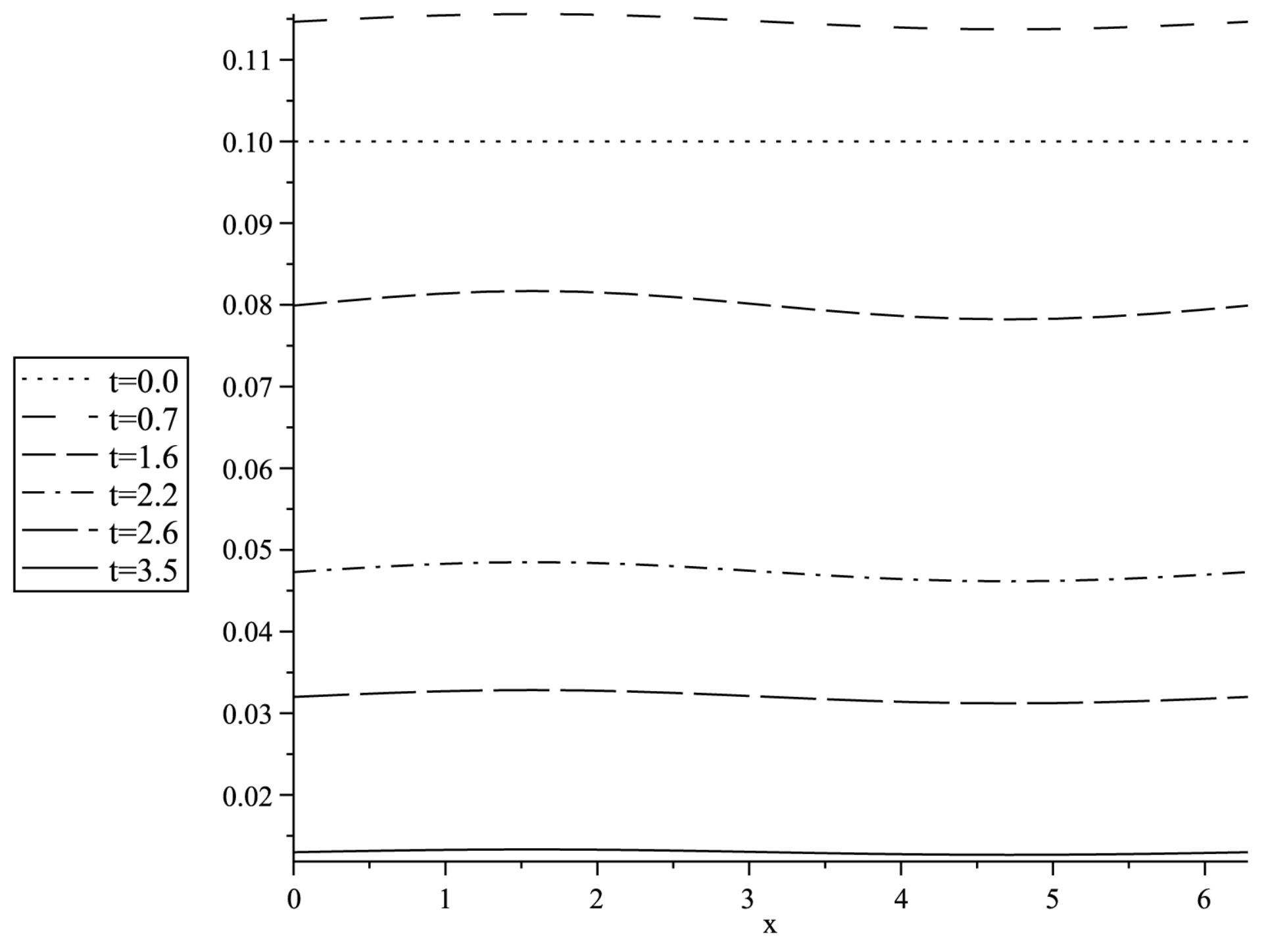}
\end{center}
\caption{Dynamics of~$E_1^1(t,x)$~at~$\gamma=1;$~$\Omega_0=0.90,~\Phi_0=0.10,~\mathcal{N}_0=0.10$}
\label{fig:edynamics}
\end{figure}

\par 
Another characteristic timepoints are~${t=t_E}$ and~${t=t_\mathcal{N}}$. These are the moments when quantities~$E_1^1$ and $\mathcal{N}$ respectively reach their maximal values, marking the period when inhomogeneity and diffusion are most significant.
\par 
It is also important to keep track of the diffusion term. We denote the moment when it approaches zero (in the sense of (\ref{def:approaching})) by~$t=\tau_\mathcal{N}.$ At later times~$t>\tau_\mathcal{N}$ the cosmology will demonstrate the same behaviour as in the case without diffusion.
\par 
Finally, one can be interested in timepoint $t=\tau_r$, where the~$\beta-$gradient becomes insignificant.
\par 
Qualitatively, most of the variables demonstrate similar behaviour for dust and radiation.
According to simulation results, the characteristic timepoints are situated in the following order with respect to each other:
\begin{equation}
0<t_E,t_\mathcal{N}<\tau_r<\tau_\mathcal{N} \leq \tau<\tau_{SB}.
\end{equation}

\par 
Numerical values for~$\tau_r,~\tau_\mathcal{N},~\tau,~\tau_{SB}$ are obtained using condition (\ref{def:approaching}).

\begin{figure}[t]
\begin{center}
\includegraphics[width=0.8\linewidth]{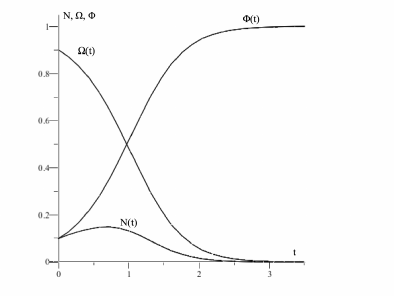}
\end{center}
\caption{Space-averaged values of~$\mathcal{N},~\Omega,~\Phi$~at~$\gamma=1$; $\Omega_0=0.90,~\Phi_0=0.10,~\mathcal{N}_0=0.10$}
\label{fig:sav}
\end{figure}

\begin{figure}[t]
\begin{center}
\includegraphics[width=0.5\linewidth]{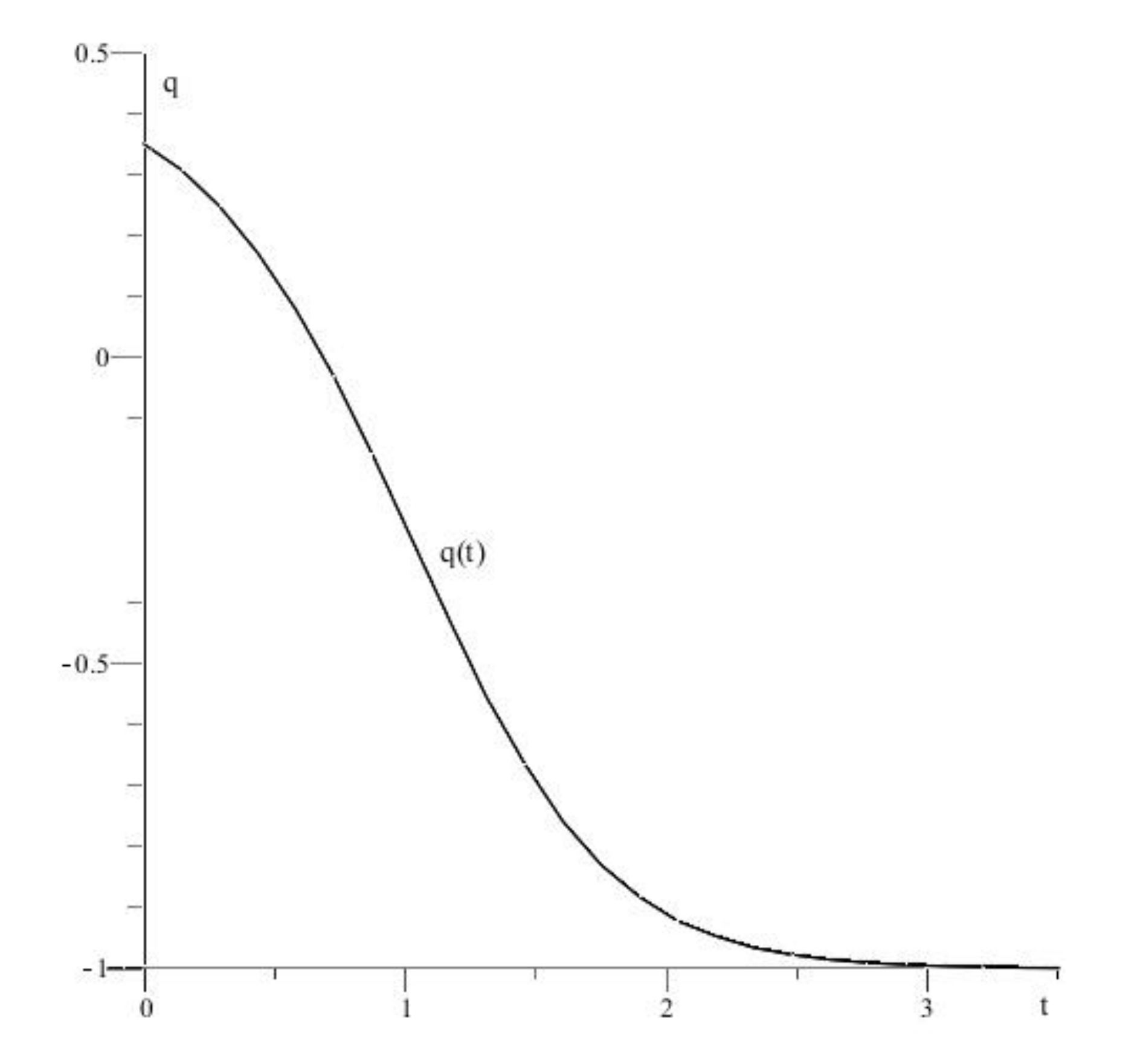}
\end{center}
\caption{Deceleration parameter~q~at~$\gamma=1$;~$\Omega_0=0.90,~\Phi_0=0.10,~\mathcal{N}_0=0.10$}
\label{fig:q}
\end{figure}

\par
The dynamics of the frame coefficient~$E_1^1$ for the dust case at initial conditions~${\Omega_0=0.90},$ $\Phi_0=\mathcal{N}_0=0.10$ is presented in Figure~\ref{fig:edynamics}. The horizontal axis shows the space coordinate~$x$, and the vertical one shows the value of~$E_1^1$. Different moments in time are depicted by lines of different kind. 

The constant initial value~$E_1^1(0,x)=E_0=0.1$ is represented by the dotted line in the figure. As the simulation starts,~$E_1^1$ becomes~$x$-dependent and experiences a slight growth; it reaches the maximum value at timepoint~$t_E=0.7$~(space-dashed line), and then decreases slowly and monotoneously. This quantity tends to zero and becomes small at the moment~$\tau_{SB}=3.5$~(solid line), when the cosmology approaches the silent boundary. For some particular initial values~(see Table~\ref{tab:phidust}, later) the maximum can disappear, and the frame coefficient decreases monotoneously towards zero.

\par 
The time evolution of space-averaged values of~$\mathcal{N},\Omega,$ and~$\Phi$ for the case of dust at the same initial conditions~${\Omega_0=0.90, \Phi_0=\mathcal{N}_0=0.10}$ is shown in Figure~\ref{fig:sav}.

\par 
The diffusion term~$\mathcal{N}$ reaches maximum value at timepoint~$t_{\mathcal{N}}=0.7$ and becomes insignificant at specific timepoint~$\tau_\mathcal{N}=2.2$. For some particular initial conditions~(see Table~\ref{tab:phidust}, later) the maximum is not observed, and then~$\mathcal{N}$ decreases monotoneously.

\begin{figure}[t]
\begin{center}
\includegraphics[width=0.8\linewidth]{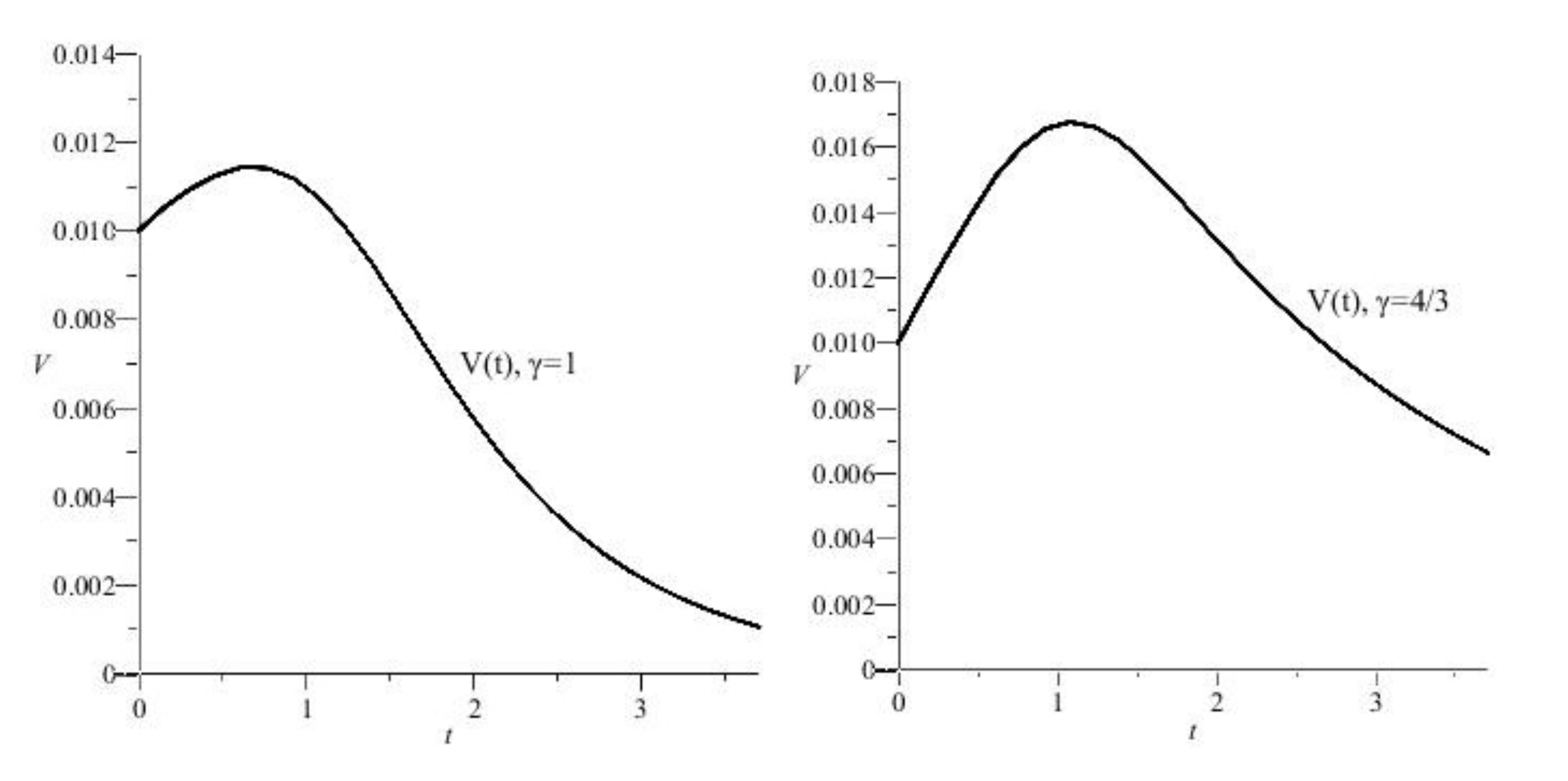}
\end{center}
\caption{Fluid velocity~$V$~at~$\gamma=1$ and~$\gamma=4/3$;~$\Omega_0=0.90,~\Phi_0=0.10,~\mathcal{N}_0=0.10$}
\label{fig:tilt}
\end{figure}

\begin{table}[t]
\begin{center}
\begin{tabular}{|ccccccccccccccccc|}
\hline $\Omega_0$ & \vline & $t_\mathcal{N}$ & \vline & $t_E$ & \vline & $\tau_r$ & \vline & $\tau_\mathcal{N}$ & \vline & $\tau$ & \vline & $\tau_{SB}$ & \vline & $E_1^1(t_E)$ & \vline & $\mathcal{N}(t_\mathcal{N})$ \\ 
\hline 0.90 & \vline & 0.7 & \vline & 0.7 & \vline & 1.6 & \vline & 2.2 & \vline & 2.6 & \vline & 3.5 & \vline & 0.11 & \vline & 0.15 \\
\hline 0.80 & \vline & 0.7 & \vline & 0.8 & \vline & 1.7 & \vline & 2.3 & \vline & 2.7 & \vline & 3.8 & \vline & 0.12 & \vline & 0.16 \\
\hline 0.70 & \vline & 0.7 & \vline & 0.8 & \vline & 1.8 & \vline & 2.3 & \vline & 2.8 & \vline & 3.9 & \vline & 0.13 & \vline & 0.18 \\
\hline
\end{tabular}
\caption{Alteration of $\Omega_0$ at $\gamma=1$.~$\Phi_0=0.10,~\mathcal{N}_0=0.10,~E_0=0.10$}
\label{tab:omegadust}
\end{center}
\end{table}

\begin{table}[t]
\begin{center}
\begin{tabular}{|ccccccccccccccc|}
\hline $\Omega_0$ & \vline & $t_E$ & \vline & $t_\mathcal{N}$ & \vline & $\tau_r$ & \vline & $\tau_\mathcal{N},\tau$ & \vline & $\tau_{SB}$ & \vline & $E_1^1(t_E)$ & \vline & $\mathcal{N}(t_\mathcal{N})$ \\ 
\hline 0.90 & \vline & 0.7 & \vline & 0.7 & \vline & 1.6 & \vline & 2.1 & \vline & 3.5 & \vline & 0.14 & \vline & 0.27 \\
\hline 0.80 & \vline & 0.8 & \vline & 0.8 & \vline & 1.6 & \vline & 2.3 & \vline & 3.7 & \vline & 0.15 & \vline & 0.29 \\
\hline 0.70 & \vline & 0.8 & \vline & 0.8 & \vline & 1.7 & \vline & 2.4 & \vline & 3.8 & \vline & 0.16 & \vline & 0.30 \\
\hline
\end{tabular}
\caption{Alteration of $\Omega_0$ at $\gamma=4/3$.~$\Phi_0=0.10,~\mathcal{N}_0=0.10,~E_0=0.10$}
\label{tab:omegarad}
\end{center}
\end{table}

\par 
Energy density~$\Omega$ and scalar potential~$\Phi$ evolve monotoneously towards their de Sitter values and approach them at the timepoint~$\tau=2.6$ (see Figure~\ref{fig:sav}).
\par 
The dependence of the space-averaged deceleration parameter~$q$ on time for the dust-filled cosmology is shown in Figure~\ref{fig:q}. Same as for~$\Omega$ and~$\Phi$, the deceleration parameter evolves monotoneously and approaches the equilibrium value~$q=1$ at~$\tau=2.6$. 

\par 
The $\beta$-gradient~$r$ is a rapidly-decaying quantity and approaches zero value at timepoint~$\tau_r$, which is significantly smaller than~$\tau$, when the cosmology approaches de Sitter state. Depending on initial conditions, $r$ can decrease monotoneously or have a subtle maximum at~${0<t<\tau_r}$.

\begin{table}[t]
\begin{center}
\begin{tabular}{|ccccccccccccccccc|}
\hline $\Phi_0$ & \vline & $t_E$ & \vline & $t_\mathcal{N}$ & \vline & $\tau_r$ & \vline & $\tau_\mathcal{N}$ & \vline & $\tau$ & \vline & $\tau_{SB}$ & \vline & $E_1^1(t_E)$ & \vline & $\mathcal{N}(t_\mathcal{N})$ \\ 
\hline 0.10 & \vline & 0.7 & \vline & 0.7 & \vline & 1.6 & \vline & 2.2 & \vline & 2.6 & \vline & 3.5 & \vline & 0.11 & \vline & 0.15 \\
\hline 0.20 & \vline & 0.1 & \vline & 0.2 & \vline & 1.2 & \vline & 1.6 & \vline & 2.2 & \vline & 3.1 & \vline & $<$0.11 & \vline & $<$0.11 \\
\hline 0.30 & \vline & -- & \vline & -- & \vline & 1.0 & \vline & 1.3 & \vline & 1.9 & \vline & 2.8 & \vline & -- & \vline & -- \\
\hline 
\end{tabular}
\caption{Alteration of $\Phi_0$ at $\gamma=1$.~$\Omega_0=0.90,~\mathcal{N}_0=0.10,~E_0=0.10$}
\label{tab:phidust}
\end{center}
\end{table}

\begin{table}[t]
\begin{center}
\begin{tabular}{|ccccccccccccccc|}
\hline $\Phi_0$ & \vline & $t_E$ & \vline & $t_\mathcal{N}$ & \vline & $\tau_r$ & \vline & $\tau_\mathcal{N},\tau$ & \vline & $\tau_{SB}$ & \vline & $E_1^1(t_E)$ & \vline & $\mathcal{N}(t_\mathcal{N})$ \\ 
\hline 0.10 & \vline & 0.7 & \vline & 0.7 & \vline & 1.6 & \vline & 2.1 & \vline & 3.5 & \vline & 0.14 & \vline & 0.27 \\
\hline 0.20 & \vline & 0.3 & \vline & 0.4 & \vline & 1.0 & \vline & 1.6 & \vline & 2.8 & \vline & 0.11 & \vline & 0.14 \\
\hline 0.30 & \vline & 0.1 & \vline & 0.2 & \vline & 0.8 & \vline & 1.4 & \vline & 2.5 & \vline & $<$0.11 & \vline & 0.11 \\
\hline 
\end{tabular}
\caption{Alteration of $\Phi_0$ at $\gamma=4/3$.~$\Omega_0=0.90,~\mathcal{N}_0=0.10,~E_0=0.10$}
\label{tab:phirad}
\end{center}
\end{table}

\par 
The fluid velocity~$V$ is a slowly decreasing quantity. In the dust case (see Figure~\ref{fig:tilt}, left) velocity tends to zero as the cosmology passes the transition stage~($t=\tau$). In case of radiation (Figure~\ref{fig:tilt}, right) the velocity decreases to some small time-independent value which is by the order of~$\epsilon.$

\section{Analysis}
\label{sec:analysis}

\subsection{Alteration of initial conditions}
\label{ssec:alteration}

\begin{table}[t]
\begin{center}
\begin{tabular}{|ccccccccccccccccc|}
\hline $\mathcal{N}_0$ & \vline & $t_E$ & \vline & $t_\mathcal{N}$ & \vline & $\tau_r$ & \vline & $\tau_\mathcal{N}$ & \vline & $\tau$ & \vline & $\tau_{SB}$ & \vline & $E_1^1(t_E)$ & \vline & $\mathcal{N}(t_\mathcal{N})$ \\ 
\hline 0.10 & \vline & 0.1 & \vline & 0.2 & \vline & 1.2 & \vline & 1.6 & \vline & 2.2 & \vline & 3.1 & \vline & $<$0.11 & \vline & $<$0.11 \\
\hline 0.20 & \vline & 0.2 & \vline & 0.3 & \vline & 1.4 & \vline & 2.0 & \vline & 2.4 & \vline & 3.3 & \vline & $<$0.11 & \vline & 0.22 \\
\hline 
\end{tabular}
\caption{Alteration of $\mathcal{N}_0$ at $\gamma=1$.~$\Omega_0=0.90,~\Phi_0=0.20,~E_0=0.10$}
\label{tab:ndust}
\end{center}
\end{table}

\begin{table}[t]
\begin{center}
\begin{tabular}{|ccccccccccccccc|}
\hline $\mathcal{N}_0$ & \vline & $t_E$ & \vline & $t_\mathcal{N}$ & \vline & $\tau_r$ & \vline & $\tau_\mathcal{N},\tau$ & \vline & $\tau_{SB}$ & \vline & $E_1^1(t_E)$ & \vline & $\mathcal{N}(t_\mathcal{N})$ \\ 
\hline 0.10 & \vline & 0.3 & \vline & 0.4 & \vline & 1.0 & \vline & 1.6 & \vline & 2.8 & \vline & 0.11 & \vline & 0.14 \\
\hline 0.20 & \vline & 0.4 & \vline & 0.5 & \vline & 1.3 & \vline & 2.0 & \vline & 3.2 & \vline & 0.11 & \vline & 0.30 \\
\hline 
\end{tabular}
\caption{Alteration of $\mathcal{N}_0$ at $\gamma=4/3$.~$\Omega_0=0.90,~\Phi_0=0.20,~E_0=0.10$}
\label{tab:nrad}
\end{center}
\end{table}

\begin{table}[t]
\begin{center}
\begin{tabular}{|lcccccccccc|}
\hline Case	& \vline & $t_E$ & \vline & $\tau_r$ & \vline & $\tau$ & \vline & $\tau_{SB}$ & \vline & $E_1^1(t_E)$ \\ 
\hline $\gamma=1$ without diffusion 		& \vline & 0.5 & \vline & 1.3 & \vline & 2.2 & \vline & 3.5 & \vline & 0.11 \\
\hline $\gamma=1$ with diffusion		& \vline & 0.7 & \vline & 1.6 & \vline & 2.6 & \vline & 3.5 & \vline & 0.11 \\
\hline $\gamma=4/3$ without diffusion		& \vline & 0.5 & \vline & 1.2 & \vline & 1.5 & \vline & 3.3 & \vline & 0.13 \\
\hline $\gamma=4/3$ with diffusion		& \vline & 0.7 & \vline & 1.6 & \vline & 2.1 & \vline & 3.5 & \vline & 0.14 \\
\hline 
\end{tabular}
\caption{Cases with and without diffusion.~$\Omega_0=0.90,~\Phi_0=0.10,~E_0=0.10$}
\label{tab:compdiffusion}
\end{center}
\end{table}

Decreasing~$\Omega_0$, while keeping the values of~$\Phi_0$ and~$\mathcal{N}_0$ fixed, leads to extension of characteristic time scales. Moreover, the maxima for~$E_1^1$ and~$\mathcal{N}$ become more significant. For the dust case, the time scales~$t_E$ and~$t_\mathcal{N}$ separate from each other having tendency~$t_E>t_\mathcal{N}$.
\par 
Results for dust are shown in Table~\ref{tab:omegadust}. For example, one can see that if the value of~$\Omega_0$ is reduced from~0.90 to~0.70, other initial constants having fixed values~${\Phi_0=\mathcal{N}_0=E_0=0.10}$, then the length of transition stage increases from~$\tau=2.6$ to~$\tau=2.8$ (approximately by 8\%). It also takes longer time for the model to approach the silent boundary: the corresponding time scale changes from~$\tau_{SB}=3.5$ to~$\tau_{SB}=3.9$ (by 11\%).
\par 
Table~\ref{tab:omegarad} shows similar data for the radiation case.

\par 
Increasing the value of~$\Phi_0$ with fixed~$\Omega_0$ and~$\mathcal{N}_0$, on the contrary, results in contraction of characteristic time scales. The maxima for~$E_1^1$ and~$\mathcal{N}$ become weaker and can disappear at significant increase in~$\Phi_0$. For dust, the time scales~$t_E$ and~$t_\mathcal{N}$ separate from each other with tendency~$t_E<t_\mathcal{N}$.

\par 
Table~\ref{tab:phidust} demonstrates these results for the case of dust. For example, if one fixes the values of initial constants to~$\Omega_0=0.90,~\mathcal{N}_0=E_0=0.10$ and increases the value of~$\Phi_0$ from~0.10 to~0.20, the length of stage with diffusion decreases from~$\tau_\mathcal{N}=2.2$ to~$\tau_\mathcal{N}=1.6$ (by~27\%), and the maximum for the diffusion term~$\mathcal{N}$ drops from~$\mathcal{N}(t_\mathcal{N})=0.15$ to~${\mathcal{N}(t_\mathcal{N})<0.11}$. If~$\Phi_0$ is increased further to~$\Phi_0=0.3$, the duration of diffusion stage decreases to~${\tau_\mathcal{N}=1.3,}$ and the maximum for the number density disappears, which is marked by dash~(-) in the table.

\par 
Corresponding data for radiation case can be found in Table~\ref{tab:phirad}.

\par 
Increasing the value of~$\mathcal{N}_0$ at fixed~$\Omega_0$ and~$\Phi_0$ leads to effects similar to those on decreasing~$\Omega_0$. The time scales extend and the maxima become more significant.
\par 
Results for dust are shown in Table~\ref{tab:ndust}, and for radiation in Table~{\ref{tab:nrad}}.

\subsection{Comparison with diffusionless models}
\label{ssec:comparison}

\begin{figure}[t]
\begin{center}
\includegraphics[width=0.6\linewidth]{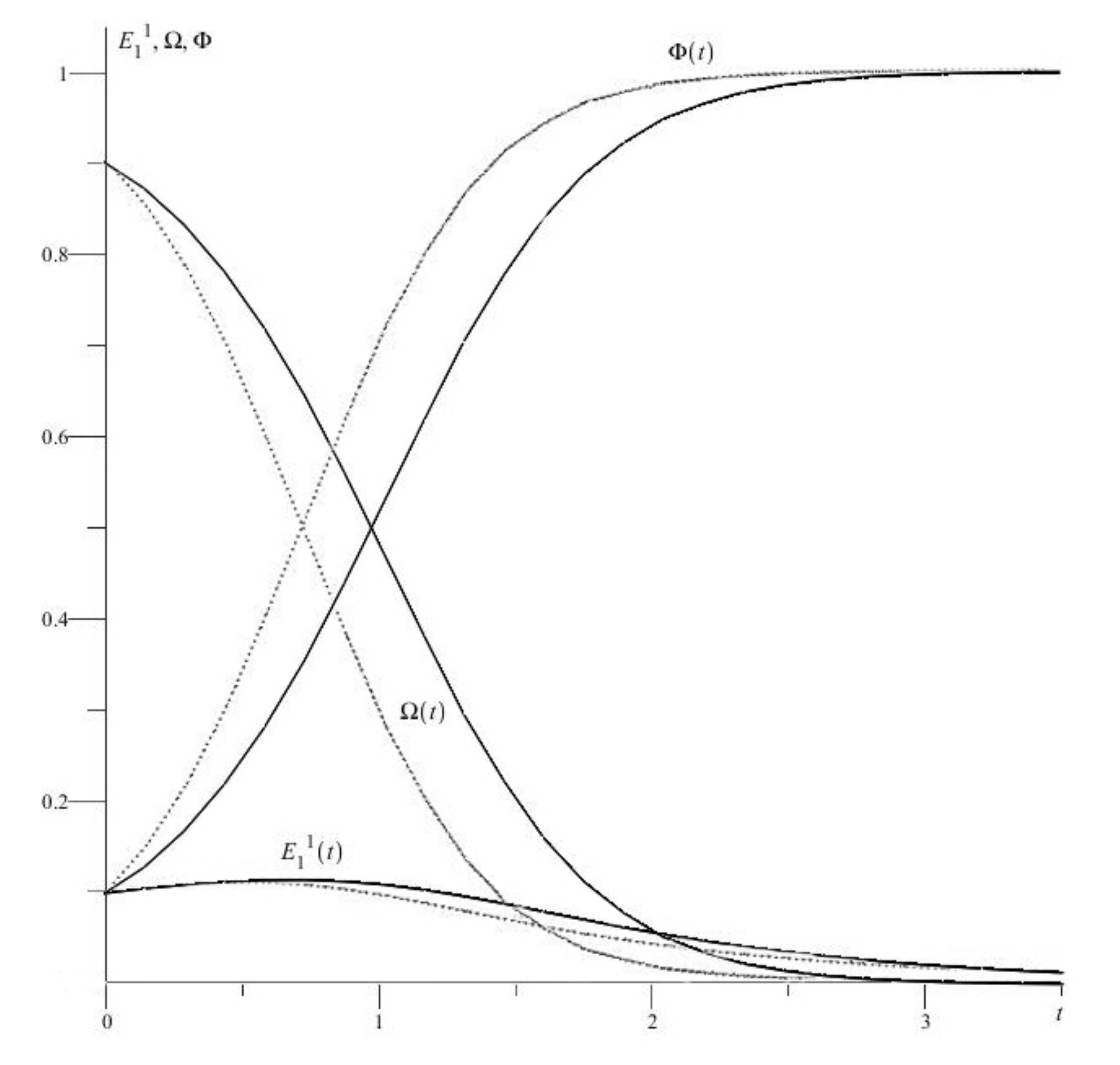}
\end{center}
\caption{Space-average values of $E_1^1, \Omega, \Phi$ at~$\gamma=1$ in the absence and presence of diffusion (dotted and solid lines respectively)}
\label{fig:diff-nodiff-dust}
\end{figure}

\begin{figure}[t]
\begin{center}
\includegraphics[width=0.6\linewidth]{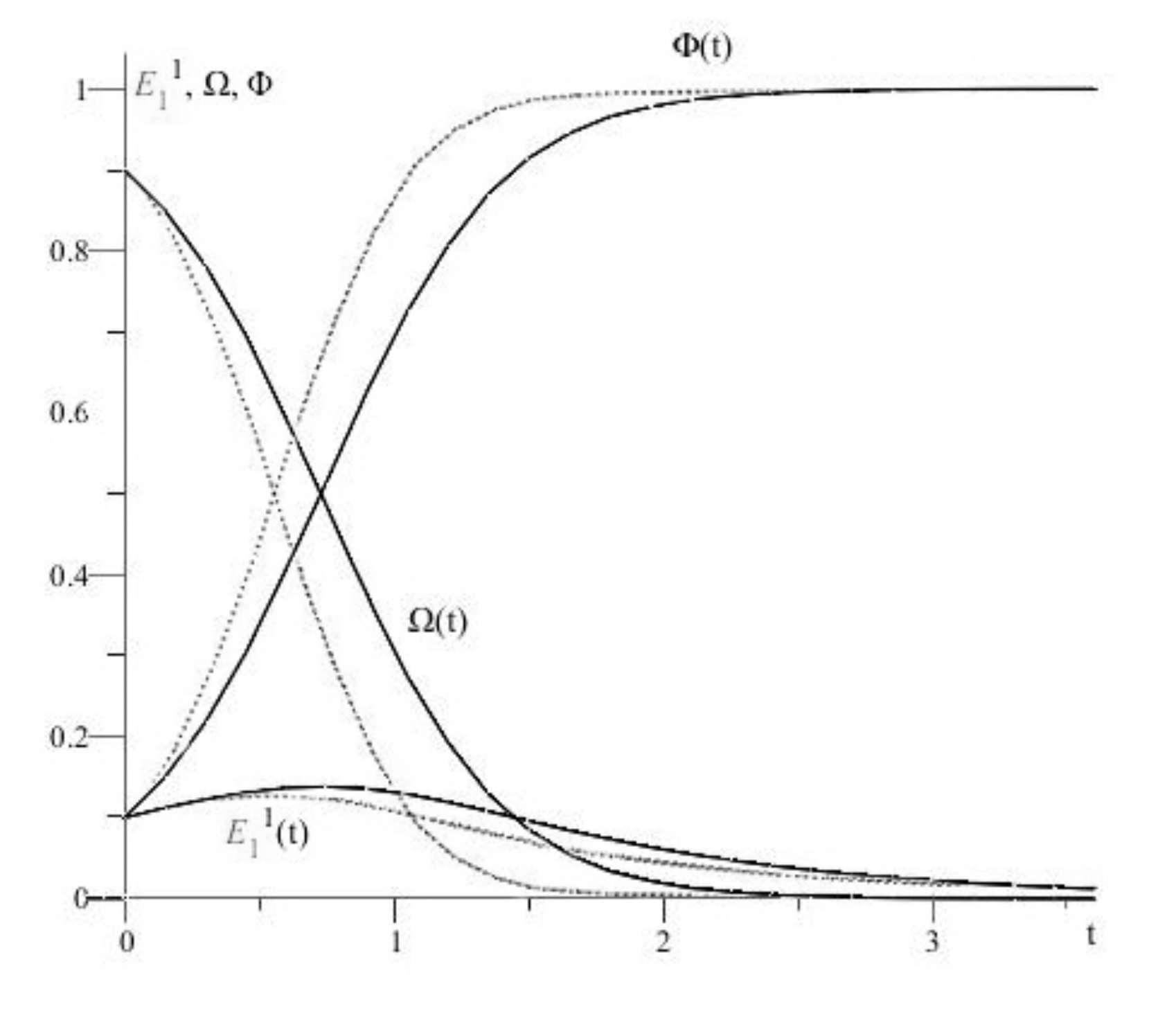}
\end{center}
\caption{Space-average values of $E_1^1, \Omega, \Phi$ at~$\gamma=4/3$ in the absence and presence of diffusion (dotted and solid lines respectively)}
\label{fig:diff-nodiff-rad}
\end{figure}

Comparison between the cases with diffusion~${(\Omega_0=0.90,~\Phi_0=0.10,~\mathcal{N}_0=0.10)}$ and without diffusion~${(\Omega_0=0.90,~\Phi_0=0.10,~\mathcal{N}\equiv 0)}$ is shown in Table~\ref{tab:compdiffusion}, Figures \ref{fig:diff-nodiff-dust} and \ref{fig:diff-nodiff-rad}.
\par 
It can be seen from Table~\ref{tab:compdiffusion} that presence of diffusion slows down the evolution of the model. For example, if we enable diffusion in the radiation-filled cosmology, then the lifetime of the~$\beta-$gradient increases from~${\tau_r=1.2}$ to~${\tau_r=1.6}$ (by about 33\%), duration of transition process increases from~$\tau=1.5$ to~$\tau=2.1$ (by 40\%), and the model approaches the silent boundary at later times:~$\tau_{SB}=3.5$ instead of previous value~$\tau_{SB}=3.3.$
\par 
In Figures~\ref{fig:diff-nodiff-dust} and~\ref{fig:diff-nodiff-rad} (for the dust- and radiation-filled model, respectively), dotted lines show the time evolution of space-averaged values of quantities~$E_1^1,~\Omega,$ and~$\Phi$ in the diffusionless case; time dynamics of the same quantites in presence of diffusion is depicted by solid lines. One can see that diffusion "stretches" the plots along the time-axis, extending the characteristic timescales.

\par 
This result is expected from the physical point of view. Under the process of cosmological diffusion, energy is transferred from the scalar field to the matter. So the energy density~$\Omega$ of matter can be predicted to decrease slower compared to the diffusionless case, which is confirmed by our numerical results.  

\subsection{Difference between Dust- and Radiation-filled models}
\label{ssec:dust_rad_difference}
\begin{enumerate}

\item 
{\bf Transition time scales.} The simulation shows that at the same initial conditions the radiation-filled model comes to the de Sitter stage~$(t=\tau)$ much faster than the dust-filled one. The radiation-filled model requires also less time to approach the silent boundary~$(t=\tau_{SB})$. This corresponds to the well-known fact (see e.\thinspace g. \cite{Groen2007}) that energy density of radiation decays faster than that of matter. Numerical results are demonstrated in Table~\ref{tab:comptimescales}.

\begin{table}[t]
\begin{center}
\begin{tabular}{|ccccccccccccc|}
\hline \multicolumn{5}{|c}{Initial constants} & \vline & \multicolumn{3}{c}{de Sitter:~$\tau$} & \vline & \multicolumn{3}{c|}{Silent Boundary:~$\tau_{SB}$} \\
\hline $\Omega_0$	& \vline & $\Phi_0$ & \vline & $\mathcal{N}_0$ & \vline & dust  & \vline & radiation & \vline & dust & \vline & radiation \\ 
\hline 0.7 & \vline & 0.1 & \vline & 0.1 & \vline & 2.8 & \vline & 2.4 & \vline & 3.9 & \vline & 3.8 \\
\hline 0.8 & \vline & 0.1 & \vline & 0.1 & \vline & 2.7 & \vline & 2.3 & \vline & 3.8 & \vline & 3.7 \\
\hline 0.9 & \vline & 0.1 & \vline & 0.1 & \vline & 2.6 & \vline & 2.1 & \vline & 3.5 & \vline & 3.5 \\
\hline 0.9 & \vline & 0.2 & \vline & 0.2 & \vline & 2.4 & \vline & 2.0 & \vline & 3.3 & \vline & 3.2 \\
\hline 0.9 & \vline & 0.2 & \vline & 0.1 & \vline & 2.2 & \vline & 1.6 & \vline & 3.1 & \vline & 2.8 \\
\hline 0.9 & \vline & 0.3 & \vline & 0.1 & \vline & 1.9 & \vline & 1.4 & \vline & 2.8 & \vline & 2.5 \\
\hline
\end{tabular}
\caption{Comparison of time scales for dust- and radiation-filled cosmologies}
\label{tab:comptimescales}
\end{center}
\end{table}

\item 
{\bf Extra stage in the dust-filled model.} In the model filled with radiation, the diffusion term~$\mathcal{N}$ becomes small at~$t=\tau$, when the model reaches the de Sitter state. Diffusion is therefore significant during the entire transition stage. However, in the cosmology filled with dust the diffusion term decays faster~(at~$t=\tau_\mathcal{N}<\tau)$. The transition process for the dust-filled cosmology has therefore a specific stage, where the model behaves like the diffusionless one. The diffusionless stage can last for~{15--30\%} of transition era.

\item 
{\bf Positions of the maxima.} In both models the timepoints~$t_E$ and~$t_\mathcal{N}$ are very close to each other. Depending on initial conditions, the time difference between these points can be extended. For the dust-filled model, all three situations are possible:~$t_E<t_\mathcal{N}$, $t_E \approx t_\mathcal{N}$, and~$t_E>t_\mathcal{N}$, while for the radiation-filled model the only tendency is~$t_E<t_\mathcal{N}.$ 

\begin{table}[t]
\begin{center}
\begin{tabular}{|ccccccccccccc|}
\hline \multicolumn{5}{|c}{Initial constants} & \vline & \multicolumn{3}{c}{$E_1^1(t_E)/E_1^1(0)$} & \vline & \multicolumn{3}{c|}{$\mathcal{N}(t_\mathcal{N}) / \mathcal{N}(0)$} \\
\hline $\Omega_0$	& \vline & $\Phi_0$ & \vline & $\mathcal{N}_0$ & \vline & dust  & \vline & radiation & \vline & dust & \vline & radiation \\ 
\hline 0.7 & \vline & 0.1 & \vline & 0.1 & \vline & 1.3 & \vline & 1.6 & \vline & 1.8 & \vline & 3.0 \\
\hline 0.8 & \vline & 0.1 & \vline & 0.1 & \vline & 1.2 & \vline & 1.5 & \vline & 1.6 & \vline & 2.9 \\
\hline 0.9 & \vline & 0.1 & \vline & 0.1 & \vline & 1.1 & \vline & 1.4 & \vline & 1.5 & \vline & 2.7 \\
\hline 0.9 & \vline & 0.2 & \vline & 0.2 & \vline & $<$1.1 & \vline & 1.1 & \vline & 1.1 & \vline & 1.5 \\
\hline 0.9 & \vline & 0.2 & \vline & 0.1 & \vline & $<$1.1 & \vline & 1.1 & \vline & $<$1.1 & \vline & 1.4 \\
\hline 0.9 & \vline & 0.3 & \vline & 0.1 & \vline & -- & \vline & $<$1.1 & \vline & -- & \vline & 1.1 \\
\hline
\end{tabular}
\caption{Comparison of maxima for dust- and radiation-filled cosmologies}
\label{tab:compmaxima}
\end{center}
\end{table}

\item 
{\bf Strength of the maxima.} Under the same initial conditions, the maxima reached by quantities~$E_1^1$ and~$\mathcal{N}$ are significantly stronger for the model filled with radiation. This is shown in Table~\ref{tab:compmaxima}. 

\item 
{\bf The velocity of the fluid.} In the dust-filled model the fluid velocity vanishes as the model passes the transition process. On the contrary, the radiation-filled model possesses time-independent velocity (which is by the order of~$\epsilon$) even when the cosmology is close to the silent boundary. This means that radiation retains velocity which becomes small during cosmological evolution but does not tend to zero. Exactly the same is observed in other cosmological models, see e.\thinspace g. \cite{Lim2006, Hervik2005, Hervik2007, Hervik2006}.

\end{enumerate}

\section{Summary}
\label{sec:summary}
The main goal was to investigate the role of the diffusion forces in governing the large-scale dynamics of an inhomogeneous anisotropic universe. One should notice that cosmological diffusion is treated to be {\it not} a fundamental interaction, but an {\it approximation} to describe the interaction between two particle systems. It was shown that under this interaction the energy is transferred from the scalar field to the matter, which slows down the process of cosmological evolution and causes the stage of accelerated expansion to arrive at later times.
\par 
In this paper one of the simplest possible diffusion models has been considered. So, some available directions of further work may be: to investigate cosmologies with more complicated geometry, include more realistic energy-matter content instead of just a perfect fluid or consider a more advanced diffusion model.
\par 
Also, one should pay attention to initial conditions introduced in Sec.\thinspace \ref{sec:ibc}. These conditions restrict the whole class of plane symmetric~$G_2$ cosmologies to a specific group we are interested in. Namely, the models passing the Robertson-Walker stage are all ever-expanding and tending to de Sitter in the asymptotic future (see e.\thinspace g. \cite{Wald1983} for homogeneous cosmologies and \cite{Lim2004a} for inhomogeneous ones). At the same time models passing far enough from Robertson-Walker equilibrium point may recollapse. This means that the sign of Hubble parameter~$H$, as well as of the expansion parameter~$\beta$, may change to the opposite at some point, so that the dimensionless variables introduced in Sec.\thinspace \ref{sec:dimensionless_variables} blow up to infinity. So, if one would like to investigate the dynamics of this group of solutions, variables~$\beta$ and~$H$ can no longer be used as normalization factors.

\bibliography{Bib/New}

\providecommand{\href}[2]{#2}\begingroup\raggedright\begin{thebibliography}{10}

\bibitem{Dudley1966}
R.~M. Dudley, {\it Lorentz-invariant markov processes in relativistic phase
  space},  {\em Ark. Mat.} {\bf 6} (1966) 241--268.

\bibitem{Risken1996}
H.~Risken, {\em The Fokker-Planck equation: methods of solution and
  applications}, vol.~18 of {\em Springer series in Synergetics}.
\newblock Springer-Verlag, Berlin, 1996.

\bibitem{Rendall2004}
A.~Rendall, {\em The Einstein-Vlasov system}.
\newblock The Einstein equations and the large scale behavior of gravitational
  fields. Birkhauser, Basel, 2004.

\bibitem{Haba2009}
Z.~Haba, {\it Relativistic diffusion},  {\em Phys. Rev. E} {\bf 79} (2009)
  021128. [arXiv:0809.1340].

\bibitem{Herrmann2009}
J.~Herrmann, {\it Diffusion in the special theory of relativity},  {\em Phys.
  Rev. E} {\bf 80} (2009) 051110. [arXiv:0903.0751].

\bibitem{Herrmann2010}
J.~Herrmann, {\it Diffusion in the general theory of relativity},  {\em Phys.
  Rev. D} {\bf 82} (2010) 024026. [arXiv:1003.3753].

\bibitem{Calogero2011}
S.~Calogero, {\it A kinetic theory of diffusion in general relativity with
  cosmological scalar field},  {\em J. Cosm. Astrop. Phys.} {\bf 11} (2011)
  016. [arXiv:1107.4973].

\bibitem{Calogero2012}
S.~Calogero, {\it Cosmological models with fluid matter undergoing velocity
  diffusion},  {\em J. Geom. Phys.} {\bf 62} (2012), no.~11 2208--2213.
  [arXiv:1202.4888].

\bibitem{Wainwright1997}
J.~A. Wainwright and G.~F.~R. Ellis, {\em Dynamical Systems in Cosmology}.
\newblock Cambridge University Press, 1997.

\bibitem{Ellis2012}
G.~F.~R. Ellis, R.~Maartens, and M.~A.~H. MacCallum, {\em Relativistic
  Cosmology}.
\newblock Cambridge University Press, 2012.

\bibitem{King1973}
A.~R. King and G.~F.~R. Ellis, {\it Tilted homogeneous cosmological models},
  {\em Commun. Math. Phys.} {\bf 31} (1973) 209.

\bibitem{Lim2004a}
W.~C. Lim, {\em Dynamics of inhomogeneous cosmologies}.
\newblock PhD thesis, University of Waterloo, 2004.
\newblock [arXiv:gr-qc/0410126].

\bibitem{Groen2007}
{\O}.~Gr{\o}n and S.~Hervik, {\em Einstein's General Theory of Relativity with
  Modern Applications in Cosmology}.
\newblock Springer, 2007.

\bibitem{Lim2004}
W.~C. Lim, H.~van Elst, C.~Uggla, and J.~A. Wainwright, {\it Asymptotic
  istropization in inhomogeneous cosmology},  {\em Phys. Rev. D} {\bf 69}
  (2004) 103507--1. [arXiv:gr-qc/0306118].

\bibitem{Lim2006}
W.~C. Lim, C.~Uggla, and J.~A. Wainwright, {\it Asymptotic silence-breaking
  singularities},  {\em Class. Quant. Grav.} {\bf 23} (2006) 2607.
  [arXiv:gr-qc/0511139].

\bibitem{Hervik2005}
S.~Hervik, R.~J. van~den Hoogen, and A.~A. Coley, {\it Future asymptotic
  behaviour of tilted bianchi models of type~IV and~VII{$_h$}},  {\em Class.
  Quant. Grav.} {\bf 22} (2005) 607. [arXiv:gr-qc/0409106].

\bibitem{Hervik2007}
S.~Hervik, R.~J. van~den Hoogen, W.~C. Lim, and A.~A. Coley, {\it Late-time
  behaviour of the tilted bianchi type~VI{$_h$} models},  {\em Class. Quant.
  Grav.} {\bf 24} (2007) 3859. [arXiv:gr-qc/0703038].

\bibitem{Hervik2006}
S.~Hervik and W.~C. Lim, {\it The late-time behaviour of vortic bianchi type
  VIII universes},  {\em Class. Quant. Grav.} {\bf 23} (2006) 3017.
  [arXiv:gr-qc/0512070].

\bibitem{Wald1983}
R.~M. Wald, {\it Asymptotic behavior of homogeneous cosmological models in
  presence of a positive cosmological constant},  {\em Phys. Rev. D} {\bf 28}
  (1983) 2118.

\end{thebibliography}\endgroup
\end{document}